\documentclass[a4paper,12pt]{article}
\usepackage{wrapfig} 
\usepackage{epsfig,amsfonts,url,stmaryrd}
\usepackage{graphicx}
\usepackage{float}
\usepackage{color}

\newcommand{\beqn}{\begin{equation}}
\newcommand{\eeqn}{\end{equation}}
\newcommand{\beqna}{\begin{eqnarray}}
\newcommand{\beqnao}{\begin{eqnarray*}}
\newcommand{\eeqna}{\end{eqnarray}}
\newcommand{\eeqnao}{\end{eqnarray*}}
\newcommand{\ba}{\begin{array}}
\newcommand{\ea}{\end{array}}

\newcommand{\unit}{\relax\ifmmode{\rm 1\>\!\!\!I}\else{$\rm 1\!I$}\fi}
\newcounter{imageCounter}

\begin{document}
\def\q{Q}

\title{Storing unsteady energy, like photovoltaically generated electric energy, as potential energy}
\date{Feb 13, 2012}
\author{ Nadja Kutz\footnote{
    email: \protect\url{nad@daytar.de}}
  }

\maketitle

\begin{abstract}
A proposal to store unsteady energy in potential energy via lifting masses with a rough quantitative overview. Some applications and methods to harvest the potential energy are also given. A focus is put on photovoltaically generated energy.
\end{abstract}

\subsection{Storing unsteady energy in potential energy}

 Electric energy generated by photovoltaical, or thermo elements - here abbreviated as photosolar or thermosolar energy or -electricity, as well as energy generated by wind is rather unstable and in some sense unsteady and  unreliable, in particular the unpredictability of weather provides still a rather big uncertainty. Thus these energy forms shall be abbreviated as ``unsteady energy''or intermittend energy. It may not in all cases be possible to level out their unstability within a smart grid. Levelling out inbalances will especially be a problem if the share of unsteady renewable energy is big in comparision to other (renewable) steady means of electricity production.

It seems to be well established that for example lifting goods in warehouses at times where there is an electricity overload may be a method to balance unsteady electrical energy within a smart grid via potential energy. Using water reservoirs may be another established method. For example Lenzen \cite{Lenzen2010} p. 547 writes:

{\em As with wind power, solar power is well suited for demands that are insensitive to temporal variations in electricity output, such as water pumping for irrigation, or vehicle charging. Matching such loads with variable solar resources can overcome restrictions related to limited capacity credit.}

These applications look however more like ``niche''- applications. That is the ``storing into potential energy'' seems not be seen as a major possibility to  make unsteady energy more reliable. So for example one finds in
{\bf Weissbuch} p. 45 under {\bf R\&D needs, opportunities and recommendations}:

{\em Electricity storage and transport: supercapacitors, supra conductors, electrolysers should be further developed.}

Lenzen writes about the storage of wind energy (p. 541):

{\em Dedicated load-leveling applications such as desalination, aluminium smelting, space and water heating, or a chargeable hybrid vehicle fleet can deal with hourly variations in wind power since they only require a certain amount of energy over a period of many hours [202,215]. For example, large-scale vehicle-to-grid technology can significantly reduce excess wind power at large wind penetration and replace a significant fraction of regulating capacity, but as Lund and Kempton [239] shows in a study for Denmark, electric vehicles would not nearly eliminate excess power and CO2 emissions, even if they had long-range battery storage.}

The DLR \cite{DLR2005} writes about an envisaged portfolio of future energy generation (p. 119):
{\em The different technologies of our portfolio contribute differently to secured power: fluctuating sources like wind and PV contribute very little, while fossil fuel plants contribute at least 90 \% of their capacity to secure power on demand.}

As one reason for this choice it is written (p.120) that:
{\em PV power is strongly fluctuating and only available during daytime. There is no contribution to secured power, but a good correlation with the usual daytime power demand peak of most countries. PV is specially suited for distributed power supply.}

In the present proposal it will be argued, that the option to store unsteady energy via potential energy should be taken more into consideration. A focus on photosolar energy will be made for brevity reasons. The here suggested examples are mainly intended for giving an outlook on the possible scope of storing via potential energy. The examples are thus in particular not exhaustive, nor exclusive. The description of the applications is however hopefully detailled enough to promote these applications as ``common knowledge'' in order to prevent that applications such as ``storing potential energy via a rod and pully system.'' are filed for being patented. The author regards the employment of patents in the renewable energy sector and especially of patents of this general kind as rather problematic for a world-wide energy turnaround.  The main intention of this article is thus to promote the idea that storing intermittend energy in potential energy may not be just a side opportunity but eventually be a major possibility to level out inbalances of intermittend energy. It is intended to encourage further investigations about the scope and applicability of storing via potential energy.

\subsection{Storing photosolar energy via potential energy}

The following examples shall illustrate what kind of applications could be used for storing intermittend energy in potential energy.

The idea here is to use some of the electric energy generated e.g. by photovoltaical elements to lift masses for energy storage. In order to reduce electricity transportation the lifting of masses is preferably in the {\em vicinity} of the location where the solar electricity is generated. This is of course not mandatory that is in certain applications it may for example for architectural reasons be necessary to transfer the electricity to a farther location. 

Typical photosolar modules are mounted on racks, these are often supplied with precast concrete ballasted footings. One possibility to realize the approach is thus for example to replace conventional module racks by  mechanical racks in which the solarmodule or a group of solar module, eventually together with some ballast can be lifted up and downward. Another realization however could also include the possibility to supply the modules with a mechanical mechanism, which is {\em detached but in the vicinity of the modules} and with which {\em arbitrary} masses can be lifted upward by using the electricity which is provided by the modules. If masses are lifted upward with solar energy then these masses acquire potential energy. The potential energy can then be converted into electrical energy whenever it is needed. Lifting masses upward can thus be seen as a possibility to store solar energy. It may be an alternative to other means of energy storage like via batteries or via compressed air energy storage (CAES) etc. It may be a supplement to smart grids.

The following calculations are intended to give a brief outlook on the feasibility of this approach.

The energy produced by a typical solar  module in Germany on a winter day extending on a surface of  $1 m^2$ per day for a $100 W_{peak}$ module could range according to the business site \cite{module2012} from 20 to 700 Wh per day. According to \cite{challenges2012} the average annual yield for the german city Hamburg is $1058 kWh/kW_{peak}$, which is about $29 Wh = 104.4 kJ$ per $100 W_{peak}$ and day. A typical $1 m^2$ size  $100 W_{peak}$ solar module weighs about 12 kg, with precast concrete ballasted footings it probably weighs about $100 kgs$ if not more. If one elevates $100 kg$ concrete ballasted modules $1 m$ above ground (via their generated electricity) one stores the potential energy of about $1kJ$, which is thus roughly about 1 percent of the averaged produced energy of such a module per day in Hamburg. 

Of course the conversion of photovoltaic energy into potential energy and then back into electricity costs. In the appendix a -eventually too idealized- back on the enveloppe calculation is done for a process in which the potential energy is converted into energy via air compression. The compressed air could for example be used for steering little wind turbines or ventilation devices. The calculation displayed that $0.985 kJ$ of the potential energy of $1 kJ$ would be maintained in this process. Converting the compressed air into electric energy via a turbine would cost additional energy. So the task is to find eventually more efficient conversion methods.  Like another method to convert the potential energy could be supplied for example by using gears and rods which steer a premium efficiency electric motor. The back of the enveloppe calculation in the appendix serves mainly as an indicator for the involved tasks and for getting an impression on energy conversion costs. Note also that we have been talking here only about energy costs due to energy conversion. The actual costs for the energy conversion devices need of course also to be taken into consideration. Likewise ``market costs'' are also playing a role. 

\begin{wrapfigure}{l}{8.5cm} 
\centering
\includegraphics[width=0.9\hsize]{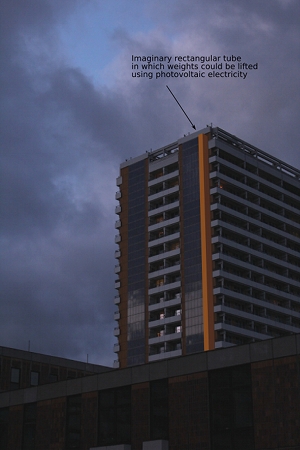}
\end{wrapfigure}

The stored potential energy can of course be scaled. For example broken basalt or concrete of about $1 m^3$ weighs about 2000 kgs \cite{simetric2012}, i.e this could give a factor of 20 for the potential energy vs the 100kg solarmodule example from above. Likewise lifting a $1 m^2$ module with  $1 m^3$ basalt underneath 10 meters up stores maximally 200kJ. Such solutions thus seem to make rather sense for applications in high buildings, where one could for example think of movable appartments (or facilities)  like with a stack of solar modules on top which are lifted along a high building (the difference to the floors can be levelled out e.g. by open stairs). The lifted masses could of course also be detached from the modules. So one could similarily imagine lifting masses in shafts which are affixed to the outside of high buildings  (see image). In both cases however one has to keep eventual noise problems into consideration. High-rising applications could also make sense at mountain sites or at remote sites (like offshore). In a landscape setting the lifting height of e.g. solar modules could be lower and for example integrated in flat valleys, in this case the up and down moving solar modules would rather look as a ``breathing solar plant'', and for the masses to be lifted, one could use excavated material or concrete.

\subsection{Conclusion}

To lift up arbitrary masses, like basalt stones, may sound as a waste of energy, however it should again be pointed out that of course a great part of the energy is regained in the process of converting the corresponding potential energy into electrical energy. Crucial points are here the efficiency of the lifting (like losses due to friction etc) and the efficiency of the conversion of the potential energy. However the losses, which are due to inefficiencies, as well as the (energy-)costs for the conversion devices have to be balanced against the possible losses due to impossibility of grid integration.

Note that many of the here proposed approaches can of course also be used for wind energy and other intermittend energy forms. For the case of wind energy one could also think about wether one converts some of the mechanical energy directly into potential energy, for example via temporarily coupling the turbine via gears to a device which lifts masses.

\subsection{Appendix}
The ideal gas law reads

$P V = n R T = const.$

In an isothermal process the energy which is put into compressing air can roughly be calculated by

$$\int_{V_A}^{V_B} P dV = n R T ln(\frac{V_B}{V_A}) = n R T ln(\frac{P_A}{P_B})$$

Hence with a molar volume of approx. $24 l/mol$ one has $1000l/(24l) \; mol = 41 mol$ for $1 m^3$, so with $R = 8.314 J/(K mol)$, one gets 

$$
41 mol *8.314 J/(mol K)*300K \simeq 100 kJ
$$

With 100 kg per $m^2$ one can thus create roughly an extra pressure of $100 kg * g/m^2 \simeq 1 kPa$, where $g = 9,80665 m/s^2$ is the standard gravity. Since 101kPa is the normal air pressure, this gives an energy equivalent of about:

$$
100 kJ * ln ((101+1)/101) = 100*9.85 * 10^{-3}kJ= 0.985kJ
$$

The calculation is intended to get a rough overview on the magnitude of possibly involved costs due to energy conversion. It is of course no complete analysis.


\begin{thebibliography}{mmmHJMNP01}



\bibitem[module2012]{module2012}
\"OKO-Energie Thomas Oberholz 
``Solarmodule zur Stromerzeugung'' 
\newblock (2012)
\newblock Consumer Information on photovoltaic modules
\newblock \url{www.oeko-energie.de/produkte/solarstrom-photovoltaik/solarmodule/index.php#04a2089a240b63601}) 

\bibitem[Lenzen2010]{Lenzen2010}
Manfred Lenzen
''Current State of Development of Electricity-Generating Technologies: A Literature Review''
\newblock (2010)
\newblock Energies 2010, 3, 462--591
\newblock \url{www.mdpi.com/1996-1073/3/3/462/pdf}

\bibitem[DLR2005]{DLR2005}
``Concentrating Solar Power for the Mediterranean Region''
\newblock German Aerospace Center (DLR)
\newblock (2005)
\newblock \url{www.dlr.de/Portaldata/1/Resources/portal.../algerien_med_csp.pdf}

\bibitem[challenges2012]{challenges2012}
Editors: Antonio Luque, Steven Hegedus
``Achievements and Challenges of Solar Electricity From PV''
\newblock (2011)
\newblock John Wiley and Sons, Ltd
\newblock Table 2.1, page 7 

\bibitem[simetric2012]{simetric2012}
simetric
``table of weights of materials''
\newblock (2012)
\newblock \url{www.simetric.co.uk/si_materials.htm} 

\bibitem[whitebook2004]{2004}
Eberhard Jochem
''Steps towards a sustainable development''
\newblock Sustainability at ETH Zürich, novatlantis
\newblock \url{http://www.novatlantis.ch/fileadmin/downloads/2000watt/Weissbuch.pdf}



\end{thebibliography}
\end{document}